\documentclass[aip, jcp, twocolumn, groupedaddress, reprint]{revtex4-1}

\usepackage{graphicx}
\usepackage{color}
\usepackage{amsmath}
\usepackage{amsfonts}
\usepackage{amssymb}
\usepackage{dcolumn}
\usepackage[T1]{fontenc}
\usepackage{hyperref}
\hypersetup{colorlinks=true, urlcolor=blue, linkcolor=blue, citecolor=blue}

\begin{document}
\title{Evidence for non-ergodicity in quiescent states of periodically sheared suspensions}
\author{K.~Julian Schrenk}
\affiliation{Department of Chemistry, University of Cambridge, Lensfield Road, Cambridge, CB2 1EW, UK}
\author{Daan Frenkel}
\email{df246@cam.ac.uk}
\affiliation{Department of Chemistry, University of Cambridge, Lensfield Road, Cambridge, CB2 1EW, UK}
\begin{abstract}
We present simulations of an equilibrium statistical-mechanics model that uniformly samples the space of quiescent states of a periodically sheared suspension.
In our simulations, we compute the structural properties of this model as a function of density.
We compare the results of our simulations with the structural data obtained in the corresponding non-equilibrium model of Cort\'e \emph{et al.}~[Nat.~Phys.~{\bf 4}, 420 (2008)].
We find that the structural properties of the non-equilibrium model are very different from those of the equilibrium model, even though the two models have exactly the same set of accessible states.
This observation shows that the dynamical protocol does not sample all quiescent states with equal probability. 
In particular, we find that, whilst  quiescent states prepared in a non-equilibrium protocol can be hyperuniform [see Phys.~Rev.~Lett.~{\bf 114}, 110602 (2015), Phys.~Rev.~Lett.~{\bf 114}, 148301 (2015), and Phys.~Rev.~Lett.~{\bf 115}, 108301 (2015)], ergodic sampling never leads to hyperuniformity.
In addition, we observe ordering phase transitions and a percolation transition in the equilibrium model that do not show up in the non-equilibrium model.
Conversely, the quiescent-to-diffusive transition in the dynamical model does not correspond to a phase transition, nor a percolation transition, in the equilibrium model. 
\end{abstract}
\maketitle
\section{\label{sec::intro}Introduction}
\begin{figure}
    \includegraphics[width=.8\columnwidth]{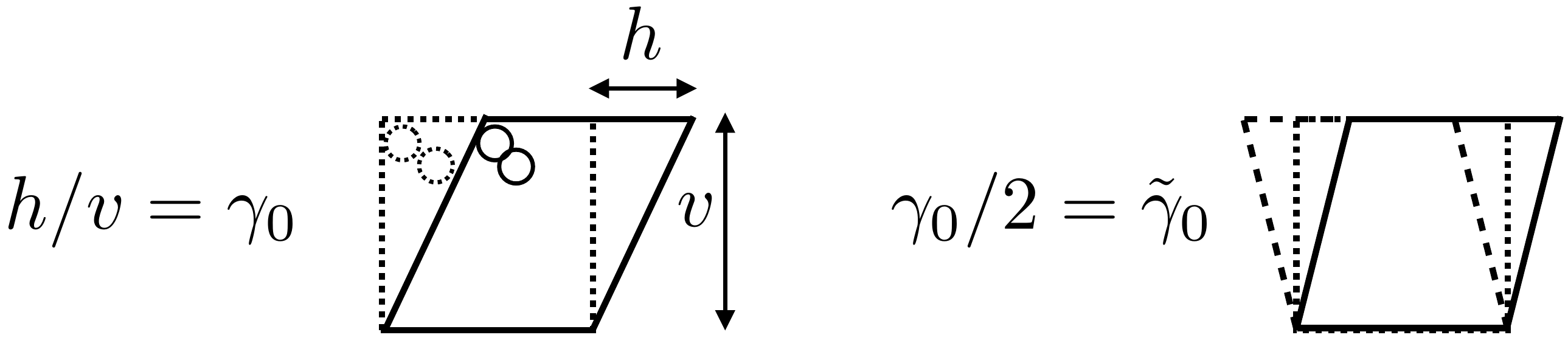}
    \caption{\label{fig::shear_transformation}
    Horizontal shear transformation of a box.
    The maximum shear amplitude $\gamma_0$ is given by the ratio of horizontal displacement $h$ over box height $v$.
    The model of Ref.~\cite{Corte08} is shown on the left, the right shows the symmetric version considered here.
    By shearing in one cycle both, to the left and the right, by half the amplitude $\tilde\gamma_0$ one obtains the same behaviour.
    }
\end{figure}
\begin{figure}
    \includegraphics[width=.8\columnwidth]{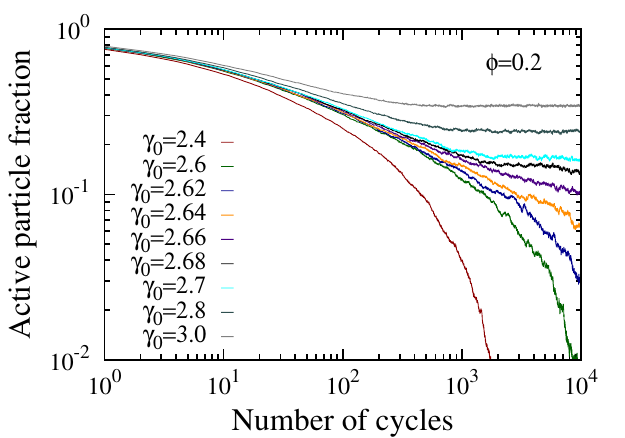}
    \caption{\label{fig::critical_gamma_averages}
    Dynamical model: fraction of active (colliding) particles in the shear cycle as function of the number of shear cycles, for different maximum shear amplitudes $\gamma_0$, at volume fraction $\phi=0.2$, for $N=1000$ particles.
    As reported in Ref.~\cite{Corte08}, the critical shear amplitude is $\gamma_{0,c}(0.2) \approx 2.66$.
    For $\gamma_0 < \gamma_{0,c}$, the activity dies out and particles arrange in quiescent states.
    Above the threshold, $\gamma_0 > \gamma_{0,c}$, the activity appears to persist indefinitely.
    At the critical point, the fraction of active particles decays as a power law with the number of cycles \cite{Corte08, Milz13}.
    }
\end{figure}
Gibbsian Statistical Mechanics is based on the assumption that the time-averaged properties of a system are equal to the ensemble averages. Although this `ergodic' hypothesis is not valid for all classical many-body systems, it is generally believed that it is correct for thermal equilibrium systems.  However, for non-equilibrium systems, there is no reason to expect ergodicity to hold.  It is, however, not straightforward to test the breakdown of ergodicity. To carry out such a test, we need to know exactly what parts of the phase space are accessible to the non-equilibrium system, and then test whether those regions  are sampled with the same frequency both in and out of equilibrium.

Here we describe simulations of a simple model system where it is possible to test the ergodicity of a non-equilibrium system.  To be more precise, we consider both a simple non-equilibrium system and an equilibrium system that has exactly the same accessible states as the non-equilibrium system. 
The non-equilibrium model that we consider was first introduced by Cort\'e \emph{et al.}~\cite{Corte08} to model the experiments by Pine \emph{et al.}~\cite{Pine05} on periodically sheared suspensions of large (non-Brownian) particles. These experiments explored the time evolution of a system of spherical plastic particles of diameter ca $0.23\,\text{mm}$, suspended in a viscous, density-matched fluid. The fluid was confined between coaxial cylinders that were sheared periodically.   Pine \emph{et al.} found that  below a critical maximum shear amplitude, after a transient period, the particle trajectories are reversible, i.e.~the particles return to their original positions at the end of each shear cycle. 

The resulting `quiescent'  arrangements of particles lack obvious geometrical order, yet they are clearly organised in such a way that particles avoid collision during a shear cycle.  Cort\'e \emph{et al.}~\cite{Corte08} introduced a simple model of the shear experiment, which starts by placing $N$ disks of diameter $\sigma$ uniformly at random in a rectangular box of area $A$ with periodic boundary conditions, giving a volume fraction of $\phi = N\pi\sigma^2/(4A)$. Each cycle consists in applying a shear transformation with maximum amplitude $\gamma_0$ in, say, horizontal ($x$-) direction and recording for each particle the number of other particles it encounters during the cycle (see Fig.~\ref{fig::shear_transformation}).
Then the box is restored to its initial shape and each particle is displaced randomly once for each encounter recorded during the cycle.
This model reproduces the threshold behaviour observed in experiments \cite{Pine05, Corte08}. Below the threshold shear amplitude, the model system reaches a quiescent state. Cort\'e \emph{et al.}~ coined the phrase ``random organisation'' to describe such states~\cite{Corte08}.

For shear amplitudes above a certain volume-fraction-dependent threshold value, particles keep colliding and random organisation into quiescent states does not occur. As an illustration, we show a typical example of the simulation results obtained with Cort\'{e}'s model:
Figure~\ref{fig::critical_gamma_averages} shows the fraction of particles that are displaced during each cycle as function of the number of cycles. At a volume fraction $\phi = 0.2$, quiescent states are observed for $\gamma_0 \lesssim 2.66$ \cite{Corte08}.
The existence of such a threshold separating reversible from non-reversible behaviour has also been reproduced in  more sophisticated Stokesian dynamics simulations \cite{Sierou01, Pine05}. However, the key non-equilibrium dynamics is already contained in the model of Ref.~\cite{Corte08}.
Cort\'{e}'s model has been argued to be in the universality class of conserved directed percolation \cite{Menon09}.

The properties of the model of Cort\'{e} \emph{et al.}~have been studied extensively.
In particular, recent studies have shown that the quiescent states that emerge at sub-critical shear amplitudes are hyperuniform, meaning that low-wavevector density fluctuations vanish in the limit that the wave-vector goes to zero~\cite{Hexner15, Tjhung15, Weijs15}.
Whilst this observation is intriguing, it is not immediately obvious whether this behaviour is a property of the quiescent states, in the sense that it would show up if we were to sample uniformly over all quiescent states, or whether the non-equilibrium dynamics does not sample the quiescent states uniformly.
In order to elucidate this question, we construct an equilibrium equivalent of the non-equilibrium model of Cort\'{e} \emph{et al.}
As we argue below, our simulations show that  different quiescent states are not sampled with equal probability by the dynamical protocol and that this `non-ergodicity' is at the origin of the observed hyperuniformity.

The remainder of this paper is organised as follows.
Section~\ref{sec::butterfly} introduces the equilibrium model that we study. In this model, all allowed states are quiescent and all quiescent states are allowed. 
In Sec.~\ref{sec::equilibrium} we report equilibrium Monte Carlo studies that sample quiescent states of the equilibrium model. 
Section~\ref{sec::hyperu} considers the nature of density fluctuations in these states.
\section{\label{sec::butterfly}Equilibrium `butterfly' model}
\begin{figure}
    \includegraphics[width=.8\columnwidth]{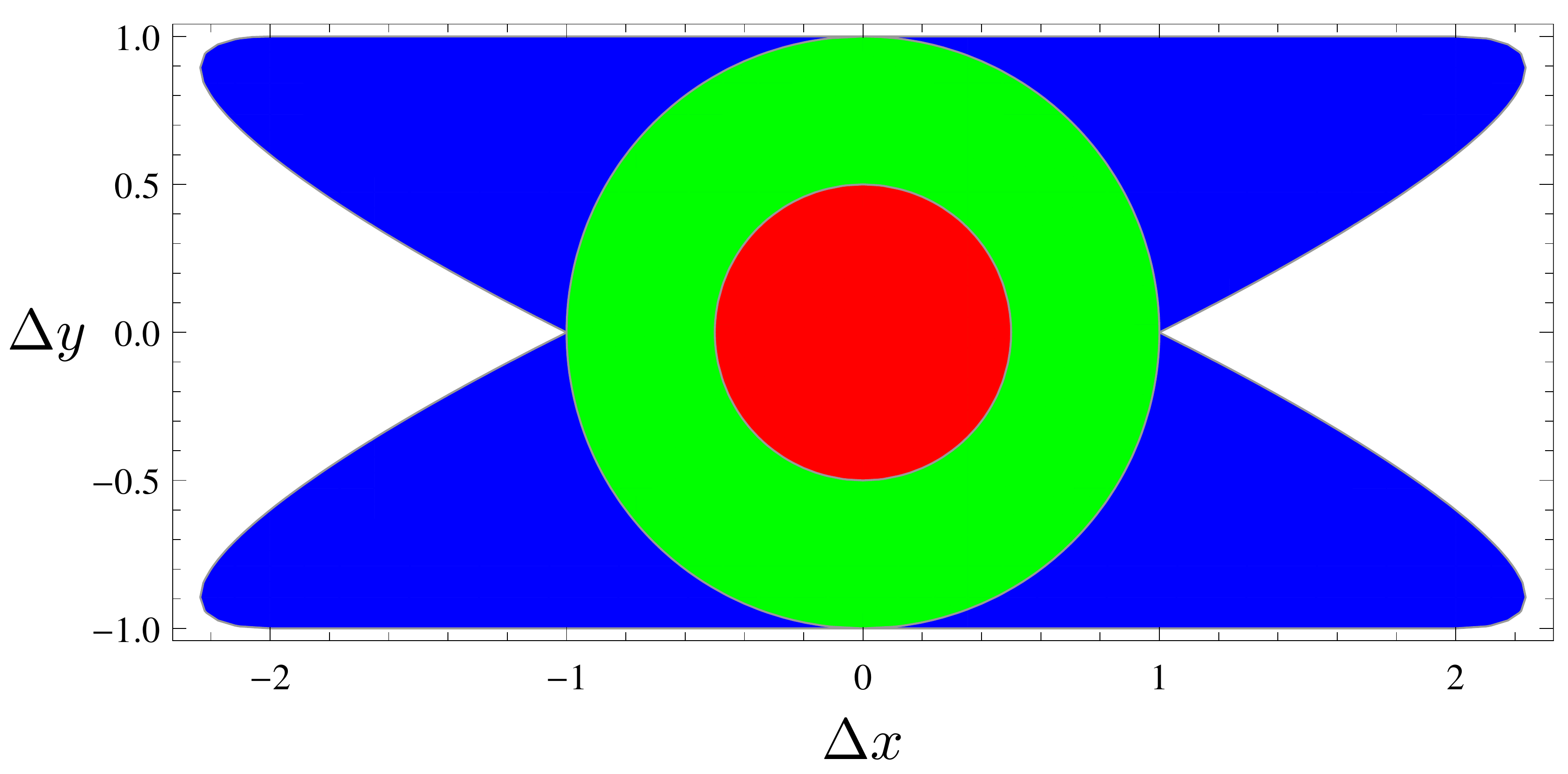}
    \caption{\label{fig::butterfly_geometry_symmetric}
        Sketch of shear butterfly geometry for unit particle diameter, $\sigma = 1$.
        The central area (red, grey) indicates the particle, the surrounding ring (green, light grey) is excluded due to hard overlap at zero shear, and the surrounding butterfly domain (blue, dark grey) is excluded as a result of one full shear cycle of maximum shear amplitude $\tilde\gamma_0 = \gamma_0/2 = 2$.
    }
\end{figure}
\begin{figure}
    \includegraphics[width=.8\columnwidth]{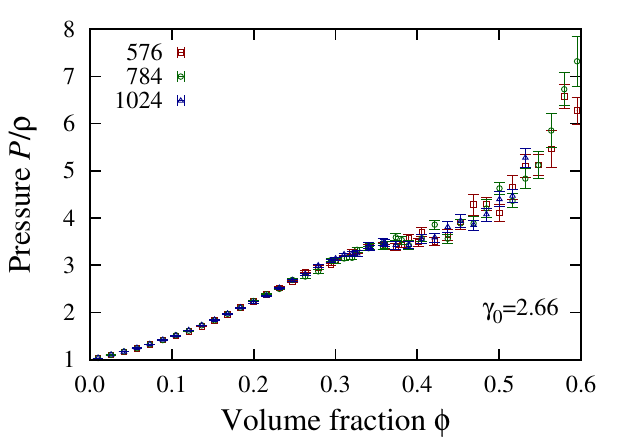}
    \caption{\label{fig::butterfly_eos_quench_full_gammaH1p33}
    Equation of state of the equilibrium butterfly fluid: The compressibility factor $P/\rho$ is shown as function of the volume fraction $\phi$ at maximum shear amplitude $\gamma_0 = 2.66$, for different number of butterflies $N$. Note that the equilibrium equation of state is smooth at ${\phi=0.2}$, the density where the transition between the quiescent and diffusive states takes place in the dynamical model.     }
\end{figure}
\begin{figure}
    \includegraphics[width=\columnwidth]{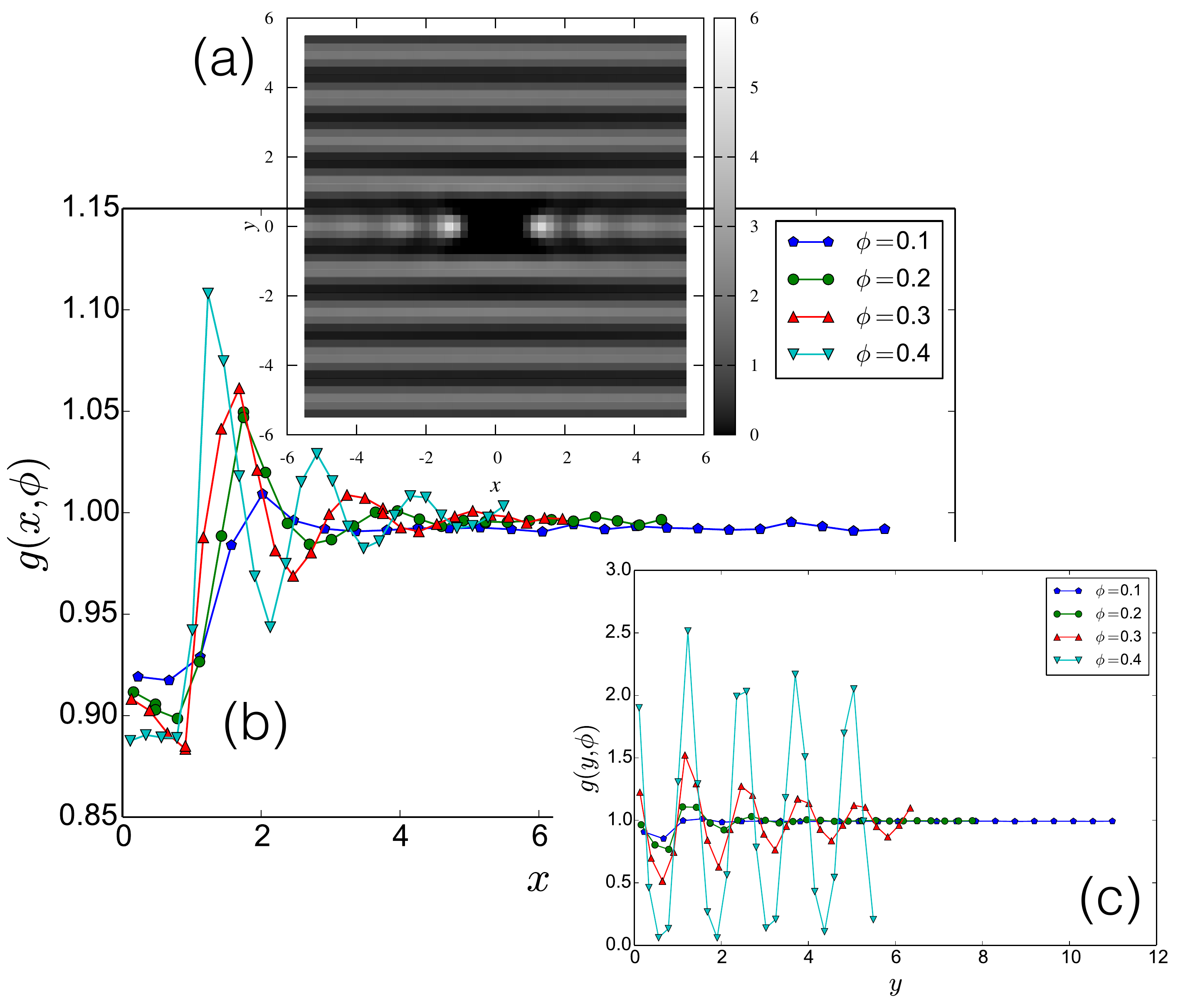}
    \caption{\label{fig::butterfly_gxy_gamma1p33}
    Pair-distance distribution ($\gamma_0=2.66$ and $N=64$): (a) shows $g(x,y)$ vs the horizontal, $x$, and vertical, $y$, centre of mass distances (see Fig.~\ref{fig::butterfly_geometry_symmetric}), at volume fraction $\phi=0.4$.
    (b) shows the vertically averaged data $g(x, \phi)$ vs $x$, for different $\phi$.
    (c) shows the horizontal average $g(y, \phi)$ vs $y$, for the same $\phi$ as in (b).
    }
\end{figure}
The dynamical shear model of Ref.~\cite{Corte08} requires the information whether or not two given disks $A$ and $B$, at positions $(x_A, y_A)^T$ and $(x_B, y_B)^T$, will overlap during the next shear cycle.
This can be determined by a straightforward geometrical argument, since the relative shear displacement of disks $A$ and $B$ only depends on their vertical distance, $\Delta y \equiv y_A - y_B$.
Parametrising the trajectories which the particles trace out in one shear cycle by $t$, such that $t=0$ corresponds to no shear and $\vert t \vert = 1$ to the maximum shear amplitude, the vector pointing from the centre of $B$ to the centre of $A$ is
\begin{equation}
    \mathbf{r} = 
    \left(\begin{array}{c} x_A(t) - x_B(t) \\ y_A(t) - y_B(t) \end{array}\right)
    = \left(\begin{array}{c} \Delta x + t \gamma_0 \Delta y \\ \Delta y \end{array}\right).
\end{equation}
We note that $\Delta x$ and $\Delta y$ are the distances before applying shear and they are therefore independent of $t$.
The particles, assumed to have the same diameter $\sigma$, collide during the shear cycle if $\vert \mathbf{r}(t) \vert \leq \sigma$ for some $-1 \leq t^* \leq 1$.
Setting $\mathbf{r}^2(t^*) = \sigma^2$ gives a quadratic equation for $t^*$ which has no solutions if $\vert \Delta y \vert > \sigma$.
However if the vertical distance $\vert \Delta y \vert \leq \sigma$, one finds
\begin{equation}
    \label{eqn::bf_t_star}
    t^* = \frac{-\Delta x \pm \sqrt{\sigma^2 - \Delta y^2}}{\gamma_0 \Delta y}.
\end{equation}
Given the requirement that $\vert t^* \vert \leq 1$, one can check for collision between disks $A$ and $B$ as follows:
\begin{itemize}
    \item If $\Delta x^2 + \Delta y^2 \leq \sigma^2$, at least one collision occurs (even without application of shear).
    \item If $\Delta y^2 > \sigma^2$, no collision occurs.
    \item If $\Delta y^2 \leq \sigma^2$, at least one collision occurs if
        $\left\vert \Delta x \pm \sqrt{\sigma^2 - \Delta y^2} \right\vert \leq \gamma_0\Delta y.$
\end{itemize}
Figure \ref{fig::butterfly_geometry_symmetric} visualises how this results in a butterfly-shaped domain around particle $A$ such that $A$ and $B$ will collide if and only if the centre of $B$ is in that domain. 

One can then place a given number of `butterflies' in a box and randomly displace the overlapping ones to simulate the dynamical model of Cort\'e \emph{et al.}~\cite{Corte08} (see Sec.~\ref{sec::intro}). Note that the butterfly model is not a hard-core model in the conventional sense, as the butterfly around $A$ excludes the {\em centre} of $B$ and conversely. Hence, overlaps of butterfly `wings' are allowed, as long as the centre of a butterfly remains outside the body of the other butterfly. 

We note that mapping from the asymmetric shear considered in Ref.~\cite{Corte08} onto the symmetric butterfly described here involves a transform of the maximum shear amplitude and of the displacement region (see Fig.~\ref{fig::shear_transformation}\cite{sm}).
\section{\label{sec::equilibrium}Equilibrium sampling of quiescent states}
Configurations in which no butterfly domain contains the centre of mass of any other particle  are by construction quiescent states.
Therefore, for fixed $N$, $\phi$, and $\gamma_0$, we can sample with equal probability among all quiescent states using a straightforward Monte Carlo algorithm \cite{Metropolis53}.
The Hamiltonian in this Monte Carlo algorithm is such that quiescent states have zero potential energy, whereas  configurations with at least one overlap, have infinite potential energy.

We measure the compressibility factor, $Z = P / \rho$ (in this paper, $k_BT=1$), where $P$ is the pressure and $\rho = N/A$ the number density, as function of the volume fraction $\phi$, for different values of the maximum shear amplitude $\gamma_0$.
The data is shown in Fig.~\ref{fig::butterfly_eos_quench_full_gammaH1p33} correspond to a system with $\gamma_0=2.66$. 
The equation of state exhibits a flat region around $\phi \approx 0.4$, which seems correlated with the density-driven transition to a striped phase, see also Fig.~\ref{fig::butterfly_gxy_gamma1p33}.
Of course,  hard disks exhibit no such striped phase\cite{sm}.  The transition to the striped phase occurs at a substantially higher volume fraction than the value $\phi_c\approx0.2$ above which quiescent states disappear in the dynamical model  (see Fig.~\ref{fig::critical_gamma_averages}): in other words, the quiescent striped phase is never observed in the dynamical model. 
We find the same behaviour for a range of $\gamma_0$ values (see summary in Fig.~\ref{fig::bf_percolation_threshold_comparison}).
Technical aspects of the pressure measurement are described in the Appendix \cite{sm}.

Figure~\ref{fig::butterfly_gxy_gamma1p33} shows the pair distribution function $g(x,y)$ at $\gamma_0=2.66$ for different volume fractions $\phi$.
We measure $g(x,y)$ by counting the number of particle pair distances, binning them according to displacement in $x$- and $y$-direction, and normalising by the number of pair distances expected if the particle distribution were homogeneous with density $\rho$.
Horizontal stripe formation is observed at $\phi=0.4$ in Fig.~\ref{fig::butterfly_gxy_gamma1p33}(a).
Vertical and horizontal averages of this data is shown in Fig.~\ref{fig::butterfly_gxy_gamma1p33}(b) and (c).
As the figure shows,  any horizontal ordering decays within a few particle diameters,  while periodic ordering persists in the vertical direction.
\section{\label{sec::hyperu}Hyperuniformity and non-ergodicity}
\begin{figure*}
    \includegraphics[width=2\columnwidth]{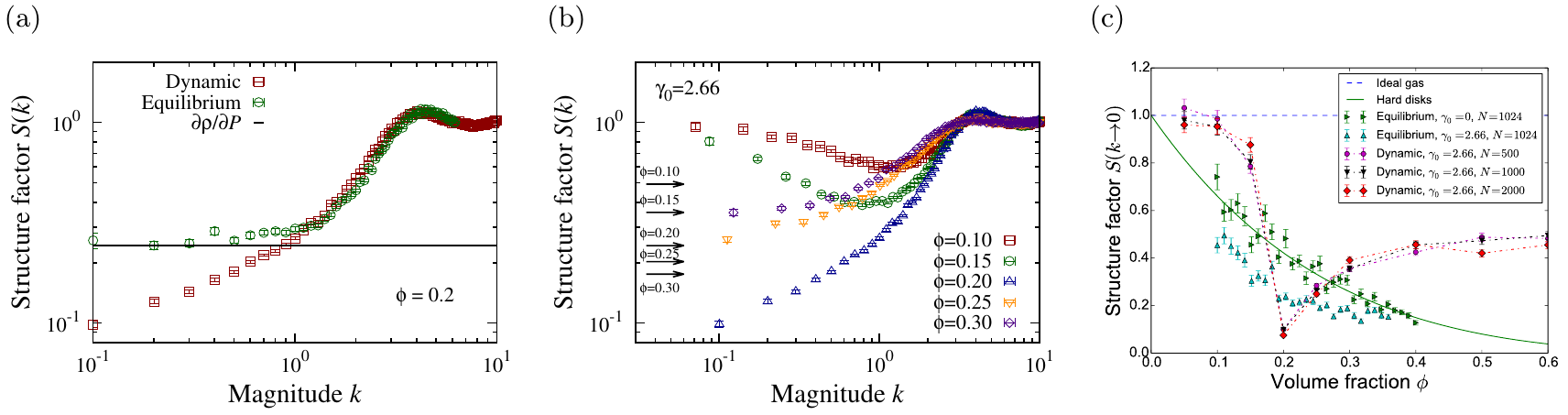}
    \caption{\label{fig::sk_eq_vs_neq}\label{fig::sk_neq_comp_sup_sub_c_Sk_to_zero}
    (a) Structure factor $S(k)$ as function of wave vector magnitude $k$ for the dynamical and the equilibrium model with $\gamma_0=2.66$ at $\phi=0.2$ (which is the critical volume fraction of the dynamical model \cite{Corte08}).
    The solid black line shows the result for $S(k\to0)$, obtained by computing $\partial\rho / \partial P$ from our independent measurement of the equation of state of the equilibrium butterfly fluid.
    (b) compares $S(k)$ for the dynamical model at different volume fractions, both, above and below $\phi_c$. 
    The solid arrows indicate the values of $S(k\to0)$ in equilibrium, obtained from $\partial\rho / \partial P$.
    (c) Structure factor low-$k$ limit $S(k\to0)$ as function of the volume fraction $\phi$ for four different models.
    For the dynamical and the equilibrium models, we plot $S(k_\text{min})$, where $k_\text{min}$ is the smallest accessible wave-vector.
    The solid green line is computed from $\partial\rho/\partial P$ with the hard disk equation of state of Ref.~\cite{Kolafa06}.
    The dashed line indicates the ideal gas result $S(k)=1$.
    }
\end{figure*}
Anomalous density fluctuations of sheared suspensions have been studied theoretically, for  the model of Ref.~\cite{Corte08} and related ones \cite{Hexner15, Tjhung15}.  Experimentally, the density fluctuations have been studied in microfluidics experiments by Weijs \emph{et al.}~\cite{Weijs15}, who also performed simulations of their experimental system. 
Over a narrow range of driving amplitudes, these studies all find evidence for hyperuniformity, i.e.~the structure factor $S(\mathbf{k})$ vanishes as $\vert\mathbf{k}\vert\to0$ \cite{Torquato03, Dreyfus15}, in some parameter regimes.
Explicitly, the structure factor $S(\mathbf{k})$, at wave vector $\mathbf{k}\neq0$, for an ensemble of $N$ points is defined by
    $S(\mathbf{k}) = \langle \rho(\mathbf{k}) \rangle / N,$
where 
    $\rho(\mathbf{k}) = \left\vert \sum_{j=1}^{N}\exp({\mathrm{i} \mathbf{k}\cdot \mathbf{r}_j}) \right\vert^2,$
with $\mathbf{r}_j$ the point positions and the average is taken over different realisations of the pattern \cite{Rosenfeld90, Frenkel13b}.
Then hyperuniformity is equivalent to
    $\lim_{\vert\mathbf{k}\vert\to0} S(\mathbf{k}) = 0,$
see e.g.~Refs.~\cite{Torquato03, Dreyfus15}.

For equilibrium systems, the long-wavelength behaviour of $S(k)$ is related to the compressibility via
    $\lim_{\vert\mathbf{k}\vert\to0} S(\mathbf{k}) = \partial \rho / \partial P,$
which, away from close packing, is finite. 
The results of Refs.~\cite{Hexner15, Tjhung15, Weijs15} are therefore indicative of the non-equilibrium nature of quiescent, randomly organised states.
However, what the existing tests cannot tell us is whether hyperuniformity is related to the average behaviour of quiescent states, or that it is due to the non-ergodic sampling of quiescent states.
As we argue below, our simulations show that it is  non-ergodicity that leads to hyperuniformity. 

Figures~\ref{fig::sk_eq_vs_neq}(a)-(b) show $S(k)$ measured for the equilibrium and dynamical models, for $\gamma_0=2.66$, at the critical volume fraction of the dynamical model, $\phi=0.2$, as well as for smaller and larger $\phi$.
We also computed $S(k=0)$ from $\partial\rho / \partial P$ (see Fig.~\ref{fig::butterfly_eos_quench_full_gammaH1p33}\cite{sm}).
Clearly, $S(k=0)$ is non-zero, and hence the equilibrium model is not hyperuniform.
By contrast the critical absorbing states of the dynamical model seem to be hyperuniform, as reported in Ref.~\cite{Hexner15} (see also Appendix \cite{sm}).
However, in the high-$k$ region $S(k)$ is similar for both models.
It seems that these short-wavelength similarities are also present in $S(\mathbf{k})$, i.e.~without radial averaging\cite{sm}.
We note that for the dynamical model at high volume fractions $\phi \gtrsim \phi_c \approx0.2$, a finite fraction of particles remains colliding, such that non-quiescent states are also found\cite{sm} and enter the computation of $S(k)$.
Figure~\ref{fig::sk_neq_comp_sup_sub_c_Sk_to_zero}(c) compares the low-$k$ region for both models with the results for hard disks at different $\phi$ and for the ideal gas.
At low $\phi$, the dynamical model behaves similarly to the ideal gas, increasing $\phi$ towards $\phi_c\approx 0.2$, $S(k\to0)$ drops and it increases again for higher $\phi > \phi_c$.
\section{\label{sec::conclusion}Conclusions}
\begin{figure}
    \includegraphics[width=.8\columnwidth]{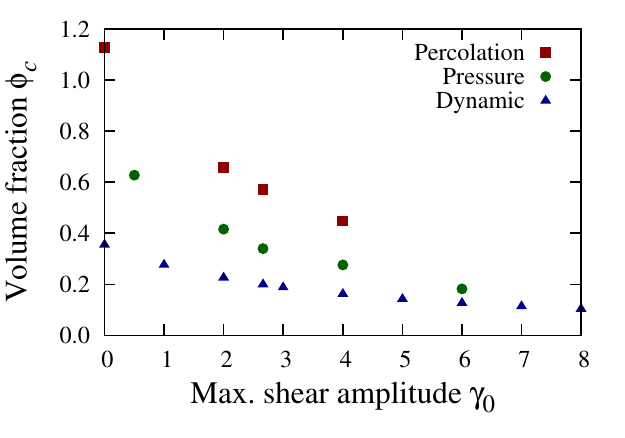}
    \caption{\label{fig::bf_percolation_threshold_comparison}
    Critical volume fractions $\phi_c$ of shear models as function of the maximum shear amplitude $\gamma_0$.
    Squares indicate the continuum percolation threshold of interpenetrable butterflies\cite{sm}.
    The location of the inflection point in the equation of state of the equilibrium butterfly fluid, as discussed in Sec.~\ref{sec::equilibrium} (see Fig.~\ref{fig::butterfly_eos_quench_full_gammaH1p33}), is indicated by circles.
    Triangles indicate the non-equilibrium critical point of the dynamical model of Ref.~\cite{Corte08} (see e.g.~Fig.~\ref{fig::critical_gamma_averages}).
    }
\end{figure}
In our study we have compared an equilibrium model that uniformly samples all quiescent states of a model for a periodically sheared suspensions of non-diffusing disks with the corresponding dynamical model. We find that, in  the dynamical model, quiescent states are sampled non-ergodically.  The differences between the dynamical and the equilibrium model are so pronounced that neither the phase transition, nor the continuum percolation threshold in the equilibrium model is related to the dynamical phase transition in the periodically sheared system. 

The different critical volume fractions are summarised in Fig.~\ref{fig::bf_percolation_threshold_comparison}.

We found that critical quiescent states are only hyperuniform if sampled by the non-equilibrium protocol of Ref.~\cite{Corte08} and not if sampled ergodically.
We note that, in principle, quiescent states can `nucleate' from diffusing states, even above the dynamical transition threshold.
It would be interesting to study the pathway for such a nucleation process because our results indicate that the nucleation `barrier' is determined by non-ergodicity, rather than by the lack of availability of quiescent states. 
\begin{acknowledgments}
This work was supported by ERC Advanced Grant 227758 (COLSTRUCTION),  EPSRC Programme Grant EP/I001352/1 and by the Swiss National Science Foundation (Grant No. P2EZP2-152188 and No. P300P2-161078).
K.J.S. acknowledges useful discussions with Nuno Ara\'ujo, Tristan Cragnolini, Daphne Klotsa, Erik Luijten, Stefano Martiniani, An{\dj}ela \v{S}ari\'c, Iskra Staneva, Jacob Stevenson, and Peter Wirnsberger. \end{acknowledgments}
\appendix
\section{\label{sec::symm_asymm_butterfly}Non-equilibrium model: transformation between symmetric and asymmetric shear}
\begin{figure}
    \begin{tabular}{l}
    (a) \\
    \includegraphics[width=\columnwidth]{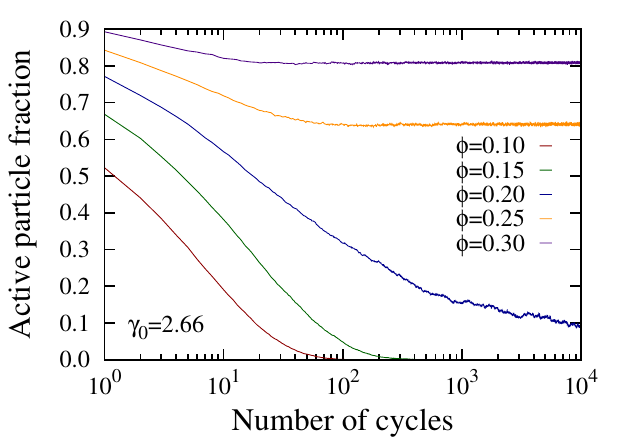} \\
    (b) \\
    \includegraphics[width=\columnwidth]{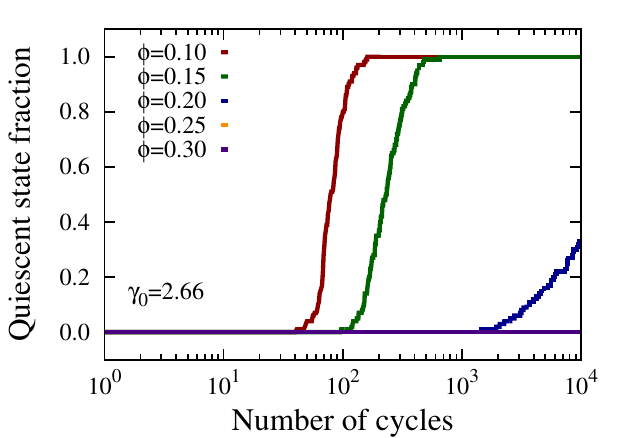}
    \end{tabular}
    \caption{\label{fig::sk_neq_comp_sup_sub_c_quiescent}
    Dynamical model with $\gamma_0=2.66$ and $N=1000$:
    (a) Fraction of active (colliding) particles as function of the number of shear cycles for different volume fractions $\phi$.
    (b) Fraction of quiescent states as function of the number of cycles for the same values of $\phi$.
    }
\end{figure}

The difference between the non-equilibrium model with symmetric and with asymmetric shear corresponds to displacing colliding disks at the end or halfway though the shear cycle, respectively (see Fig.~\ref{fig::shear_transformation}).
In the asymmetric model of Ref.~\cite{Corte08}, each displacement is sampled uniformly at random from a square of side length $\sigma$ centred at the position of the disk:
\begin{equation}
    \begin{aligned}
        y_\text{new} - y_\text{old} &= \sigma/2\left\{1-2\,\text{uni}[0,1)\right\} \\
        x_\text{new} - x_\text{old} &= \sigma/2\left\{1-2\,\text{uni}[0,1)\right\},
    \end{aligned}
\end{equation}
where $\text{uni}[0,1)$ is a uniform random number.
To simulate the same behaviour with the symmetric butterfly described here (see Fig.~\ref{fig::butterfly_geometry_symmetric}), one can first sample displacements as above and then apply a shear transformation:
\begin{equation}
    \begin{aligned}
        y_\text{new} - y_\text{old} &= \sigma/2\left\{1-2\,\text{uni}[0,1)\right\} \\
        x_\text{new} - x_\text{old} &= \sigma/2\left\{1-2\,\text{uni}[0,1)\right\} + \tilde\gamma_0 (y_\text{new}- y_\text{old}).
    \end{aligned}
\end{equation}
Figure~\ref{fig::sk_neq_comp_sup_sub_c_quiescent} shows the fraction of colliding particles and of quiescent states in the dynamical model with $\gamma_0=2.66$, for different volume fractions $\phi$.
\section{\label{sec::pressure_details}Pressure measurement methods and benchmark}
\begin{figure}
    \includegraphics[width=\columnwidth]{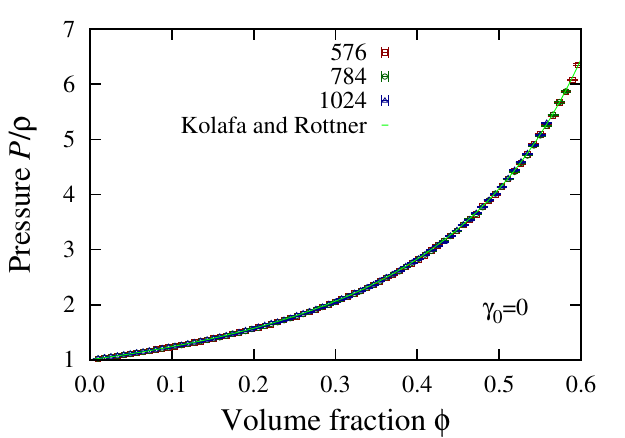}
    \caption{\label{fig::eos_disk_gamma0_benchmark}
    Pressure measurement benchmark: $P / \rho$ as function of volume fraction $\phi$ for hard disk fluids of different numbers of disks.
    Symbols refer to measurements with the method of Ref.~\cite{Eppenga84}, the solid green line shows the equation of state of Kolafa and Rottner \cite{Kolafa06}.
    }
\end{figure}

While the equation of state of hard disks (corresponding to equilibrium butterflies with $\gamma_0=0$) has been determined numerically to high precision \cite{Kolafa06}, no such results are available for $\gamma_0>0$.
To obtain the equation of state of the equilibrium butterfly fluid, we use the virtual trial volume change method of Ref.~\cite{Eppenga84}.
This method allows to measure the pressure in Monte Carlo simulations at fixed number of particles $N$, volume $V$, and temperature $T$.
The procedure is to measure the probability density $P_1(\Delta\rho)$ of a virtual density change $\Delta\rho$, such that a given particle will have its first overlap with any neighbour if the density is changed by $\Delta\rho$.
For small $\Delta\rho$, one expects for the probability density \cite{Eppenga84}:
\begin{equation}
    \label{eqn::EF_density}
    P_1(\Delta\rho) = \alpha \exp(-\alpha\Delta\rho),
\end{equation}
such that the probability is
\begin{equation}
    \text{prob}_1(\Delta\rho) = 1 - \exp(-\alpha\Delta\rho).
\end{equation}
Measuring $\alpha$ in this way allows to determine the compressibility factor:
\begin{equation}
    Z = 1 + \alpha \rho / 2.
\end{equation}
To benchmark our implementation of the pressure measurement method of Ref.~\cite{Eppenga84}, we compared our results at $\gamma_0=0$ with the hard disk equation of state put forward in Ref.~\cite{Kolafa06}, as shown in Fig.~\ref{fig::eos_disk_gamma0_benchmark}.

\begin{figure}
    \begin{tabular}{l}
    (a) \\ 
    \includegraphics[width=\columnwidth]{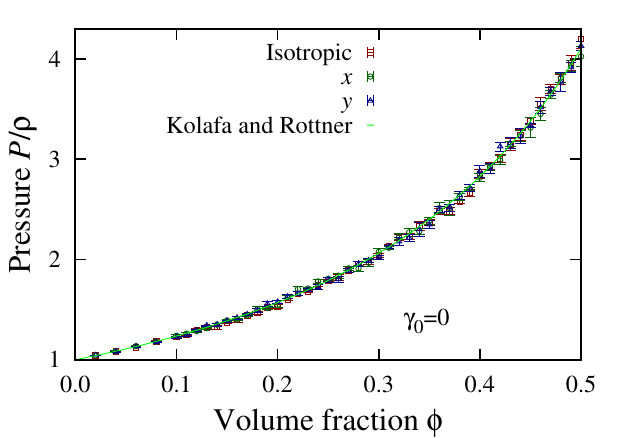} \\
    (b) \\
    \includegraphics[width=\columnwidth]{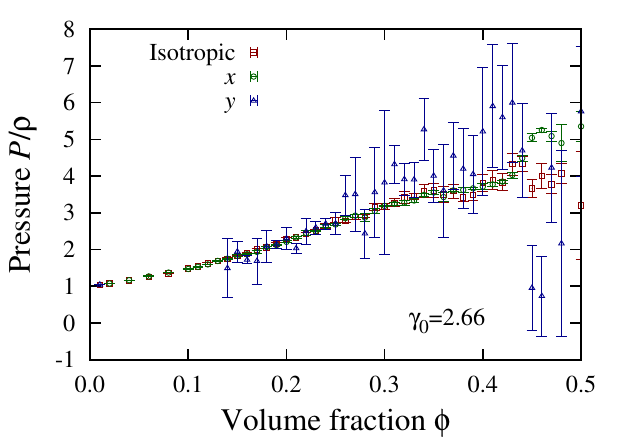} \\
    \end{tabular}
    \caption{\label{fig::butterfly_stress_anisotropy_p1}
    Compressibility factor $P/\rho$ of the equilibrium butterfly fluid as function of the volume fraction $\phi$, for (a) $\gamma_0 = 0$ (hard disks) and (b) $\gamma_0 = 2.66$, with $N=125$. The pressure was measured proposing isotropic trial volume changes (squares), volume changes only in $x$-direction (circles), and only in $y$-direction (triangles).
    }
\end{figure}

To check if the pressure (stress) is isotropic, we considered also virtual density changes due to resizing the simulation box in $x$- and $y$-direction only, respectively.
The results were consistent with the ones obtained by isotropic virtual compression, at least before the striped phase appears, see Fig.~\ref{fig::butterfly_stress_anisotropy_p1} for examples.
Due to the butterfly geometry one expects that for volume changes in $x$-direction only compression can result in overlaps, while for volume changes in $y$-direction both, compression and expansion, can yield overlaps and will contribute to the pressure measurement.
In that case we measured the probability density in Eq.~(\ref{eqn::EF_density}) for both, $\Delta\rho\to0^+$ and $\Delta\rho\to0^-$ and computed the compressibility factor as:
\begin{equation}
    Z = 1 + (\alpha^+ - \alpha^-) \rho / 2.
\end{equation}
To extract $\alpha$ from the behaviour of $\log P_1(\vert\Delta\rho\vert)$ in Eq.~(\ref{eqn::EF_density}), it was useful to consider linear regression using powers of $\Delta\rho$ as basis functions \cite{Bishop09}.
For $\Delta\rho\to0^+$, a robust choice was a model of the form $a_0 + a_1 \Delta\rho + a_2 (\Delta\rho)^3$ and for $\Delta\rho\to0^-$ of the form $a_0 + a_1 \vert\Delta\rho\vert + a_2 \vert\Delta\rho\vert^{1.5}$.
\section{Structure factor}
\begin{figure}
    \begin{tabular}{l}
        (a) \\
        \includegraphics[width=0.9\columnwidth]{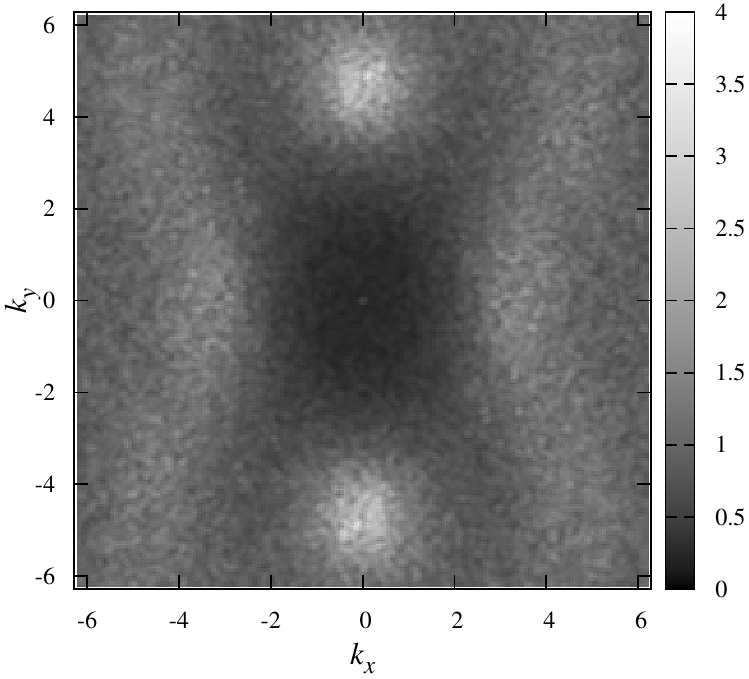} \\
        (b) \\
        \includegraphics[width=0.9\columnwidth]{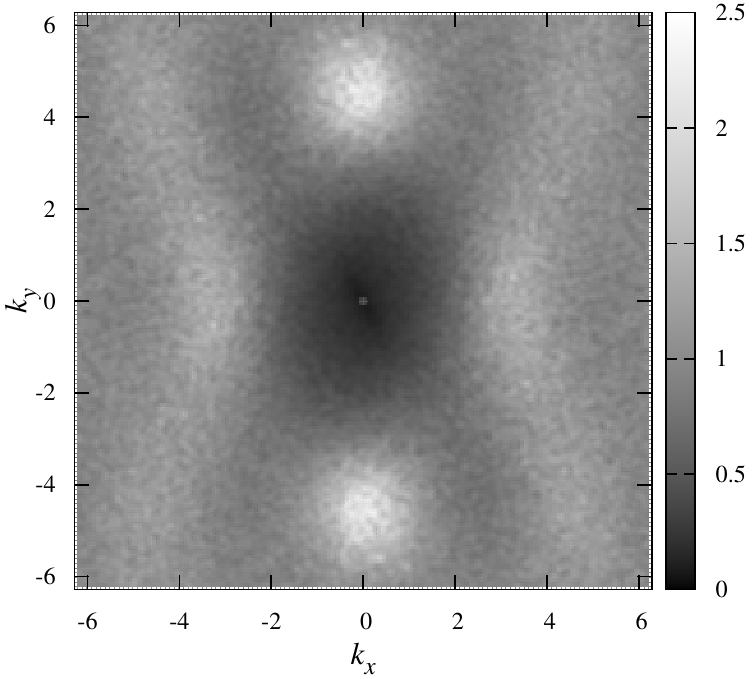} \\
        (c) \\
        \includegraphics[width=0.9\columnwidth]{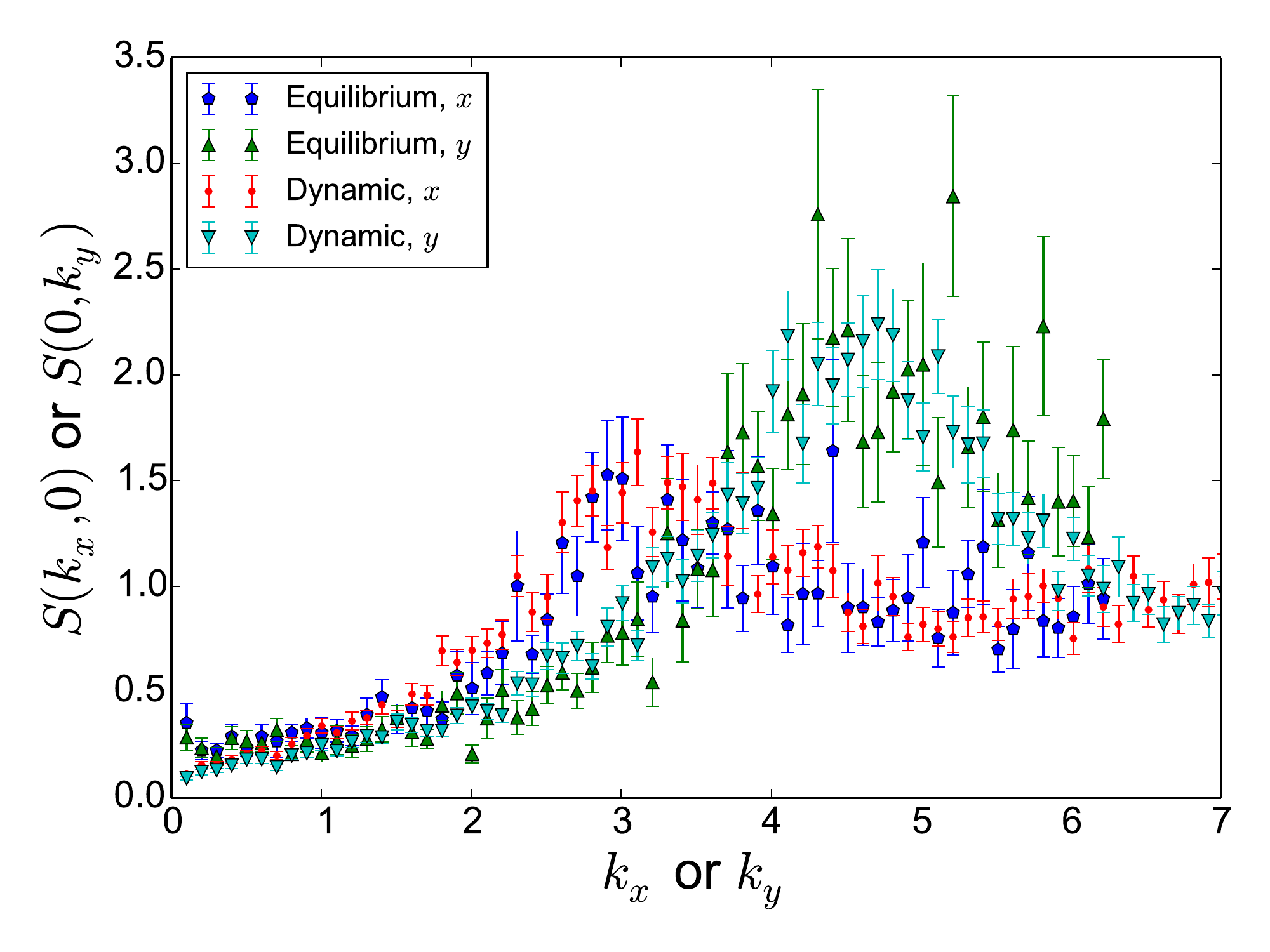} \\
    \end{tabular}
    \caption{\label{sk_eq_vs_neq_fullxy_eq_non_eq_comp}
    Structure factor $S(\mathbf{k})$ for (a) the equilibrium model, and (b) the dynamical model, for the same parameters as in Fig.~\ref{fig::sk_eq_vs_neq}.
    (c) shows $S(k_x, 0)$ and $S(0,k_y)$ as function of the wave vector magnitude $k_{x,y}$.
    }
\end{figure}

Figure~\ref{sk_eq_vs_neq_fullxy_eq_non_eq_comp} suggests that short-wavelength similarities between the dynamical and equilibrium models are also present in $S(\mathbf{k})$, i.e.~without radial averaging.
It seems that for small $k$, $S(k_x)_\text{Dynamic} \approx S(k_y)_\text{Dynamic} \to 0$ and $S(k_x)_\text{Equilibrium} \approx S(k_y)_\text{Equilibrium} > 0$, while for larger $k$, $S(k_x)_\text{Dynamic} \approx S(k_x)_\text{Equilibrium}$ and $S(k_y)_\text{Dynamic} \approx S(k_y)_\text{Equilibrium}$.

\begin{figure}
    \includegraphics[width=\columnwidth]{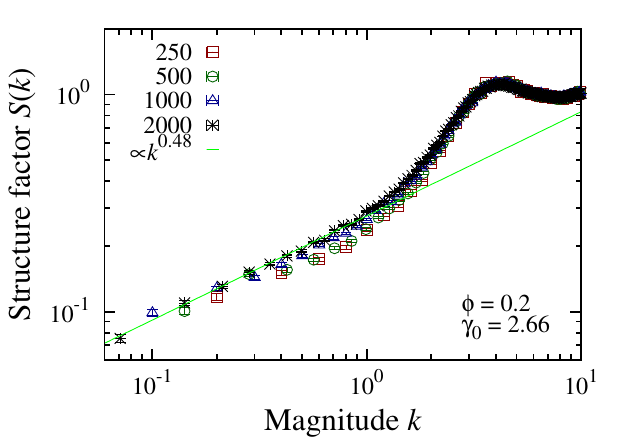}
    \caption{\label{fig::sk_non_eq_power_check}
    Structure factor $S(k)$ as function of the wave vector magnitude $k$, for the non-equilibrium model at $\gamma_0 = 2.66$ and $\phi = \phi_c = 0.2$, for different $N$.
    The solid line is a guide to the eye $\propto k^{0.48}$.
    }
\end{figure}

Hexner and Levine \cite{Hexner15} report that for the critical Conserved Lattice Gas model the structure factor behaves as $S(k)\sim k^{0.45}$.
Figure~\ref{fig::sk_non_eq_power_check} shows our data for the critical non-equilibrium model of Ref.~\cite{Corte08}.
The data seems to be compatible with a power law $S(k) \sim k^{0.48}$ in the approximate range $0.1 \lesssim k \lesssim 1$.
Given the challenges involved in determining such an exponent precisely, our result appears to be compatible with Ref.~\cite{Hexner15}.
However we can not exclude that the behaviour of $S(k)$ might be different for smaller values of $k$ than the ones accessible in our simulations.
We note that we computed $S(k)$ directly from its definition as given in Sec.~\ref{sec::hyperu} \cite{Frenkel13b}.
\section{\label{sec::percolation}\label{sec::percolation_details}Continuum percolation}
\begin{figure}
    \includegraphics[width=\columnwidth]{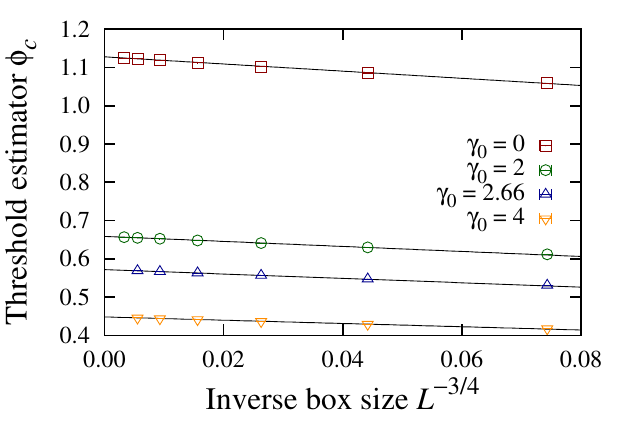}
    \caption{\label{fig::bf_percolation_jump_time}
    Continuum percolation transition: position $\phi_c$ of largest change in the order parameter as function of the inverse box size $L^{-3/4}$, for different values of the maximum shear amplitude $\gamma_0$.
    For $\gamma_0=0$, i.e.~overlapping disks, our result is compatible with the literature value: $\phi_c(L\to\infty) = 1.12808737(6)$ \cite{Mertens12}.
    }
\end{figure}

To study the emergence of non-quiescent states, we investigate the continuum percolation properties of shear butterflies (see Sec.~\ref{sec::butterfly}).
Consider a square box of side length $L$ with periodic boundary conditions.
Starting from an empty box, butterflies are placed one-by-one at positions sampled uniformly at random.
Two particles are considered to be part of the same cluster if the centre of mass of one of them is within the butterfly domain of the other.
The size of a cluster is given by the number of butterflies it contains.
We note that since disks can overlap for continuum percolation, the volume fraction is linear in the number of butterflies $N$ and it is not bounded by unity.
The order parameter of the percolation transition is the fraction of butterflies in the box which are part of the largest cluster \cite{Borgs01}.
To find the percolation threshold (i.e.~the lowest value of $\phi$ at which the probability to find a percolating cluster is non-zero), we measure for each realisation the position $\phi_c$ of the largest change in the order parameter \cite{Nagler11} due to insertion of a single butterfly.
Finally the measurements of $\phi_c$ are averaged over all sampled realisations.
To obtain percolation thresholds for the limit of infinite butterfly swarm size, we extrapolate the data according to the scaling behaviour expected from percolation theory \cite{Borgs01}:
\begin{equation}
    \phi_c(L) \sim \phi_c(\infty) + \text{const}/L^{3/4},
\end{equation}
see Fig.~\ref{fig::bf_percolation_jump_time}.
As expected,  the percolation threshold decreases with increasing maximum shear amplitude $\gamma_0$ (see the summary in Fig.~\ref{fig::bf_percolation_threshold_comparison}).
As is clear from Fig.~\ref{fig::bf_percolation_jump_time}, the percolation threshold in the equilibrium model (e.g. $\phi\approx 0.58$ for $\gamma_0=2.66$,  is not related to the onset of diffusive behaviour in the dynamical model ($\phi = 0.2$). 

To simulate continuum percolation of butterflies, we use a cell algorithm similar to Refs.~\cite{Newman00, Li09, Mertens12} to keep track of shape overlaps.
For applying cell lists, it is useful to know the butterfly wingspan.
Using Eq.~(\ref{eqn::bf_t_star}), setting $t^*=1$ and finding the maximum of $\Delta x$, gives the wingspan $w(\gamma_0, \sigma)$:
\begin{equation}
    w(\gamma_0, \sigma) = \frac{2\sigma(1+\gamma_0^2\sigma)}{\sqrt{1+\gamma_0^2}}.
\end{equation}

\begin{figure}
    \includegraphics[width=\columnwidth]{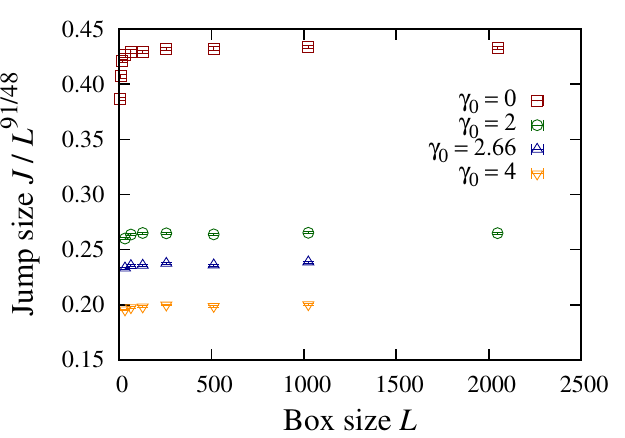}
    \caption{\label{fig::bf_percolation_jump_size}
    Percolation cluster fractal dimension: rescaled largest jump in the maximum cluster size $J/L^{91/48}$ as function of the box length $L$, for different values of the maximum shear amplitude $\gamma_0$.
    The fractal dimension is consistent with the rigorously known critical exponents of two-dimensional percolation \cite{Smirnov01b}.
    }
\end{figure}

As a further sanity check of our percolation simulations we consider the finite size behaviour of the largest change in the maximum cluster size (jump) $J$.
Percolation theory predicts that this is determined by the fractal dimension:
\begin{equation}
    J \sim L^{2 - \beta / \nu},
\end{equation}
where $2-\beta/\nu = 91/48$ is the fractal dimension \cite{Borgs01, Smirnov01b}.
Figure~\ref{fig::bf_percolation_jump_size} shows that this scaling behaviour is indeed observed for large box sizes $L$.

\begin{figure}
    \includegraphics[width=\columnwidth]{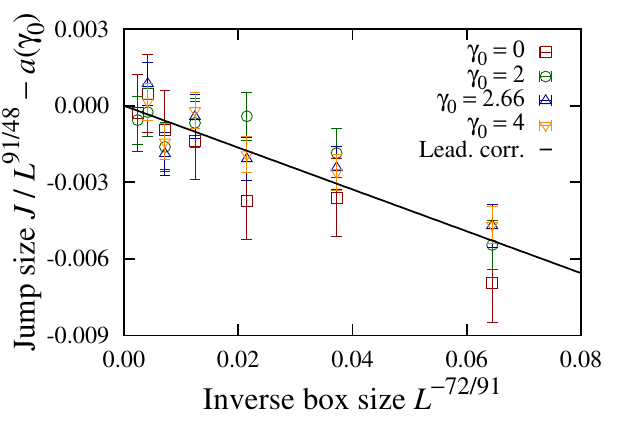}
    \caption{\label{fig::bf_percolation_jump_size_corrections_to_scaling}
    Leading corrections to scaling: rescaled and shifted jump size
    $J / L^{91/48} - a(\gamma_0)$
    as function of the inverse box size
    $L^{-72/91}$.
    The data seems to be compatible with the leading correction-to-scaling exponent proposed in Ref.~\cite{Ziff09}.
    The solid black line is a linear guide to the eye.
    }
\end{figure}

The leading correction to scaling for the jump size is expected to be of the form
\begin{equation}
    J \sim L^{2-\beta/\nu}(a + b L^{-\Omega}),
\end{equation}
where $a$ and $b$ are constants and $\Omega$ is the leading correction to scaling exponent.
In 2011, Ziff argued, based on results from conformal field theory, that for two-dimensional percolation $\Omega = 72/91$ \cite{Ziff09}.
This was supported by simulation data for lattice percolation.
Analysing the data of Fig.~\ref{fig::bf_percolation_jump_size}, we check this prediction for continuum percolation of butterflies.
Figure~\ref{fig::bf_percolation_jump_size_corrections_to_scaling} shows the rescaled and shifted jump size
$J / L^{91/48} - a(\gamma_0)$ as function of the inverse box size $L^{-72/91}$.
The data is compatible with a linear relationship.
\bibliography{butterflies}
\end{document}